\documentclass[10pt]{article}
\usepackage{graphicx}% Include figure files
\usepackage{epstopdf}
\usepackage{amstext,amsfonts,amsbsy,amssymb,bbm}
\usepackage{array,multirow}
\usepackage[pageshow]{supertabular}
\usepackage{pdfsync}
\usepackage{a4wide}
\usepackage{times,color}
\newcommand{\Ker}{\mathop{\mbox{Ker}}}

\def\Im{\mathop{\rm Im}\nolimits}

\def\Tr{\mathop{\rm Tr}\nolimits}

\newcommand {\be}[1]{\begin{eqnarray} \mbox{$\label{#1}$}  }
\renewcommand {\be}[1]{
    \begin{eqnarray} \mbox{$\label{#1}$}  }
\newcommand{\ee}{\end{eqnarray}}

\newcommand{\pref}[1]{(\ref{#1})}

\newcommand{\iden}{\mathbbm{1}}

\newcommand\ie {{\it i.e.}, }

\newcommand\etal {{\it et. al. }}

\newcommand{\noi}{\noindent}

\newcommand{\bs}{\boldsymbol}

\newcommand{\pdd}[2]{{\partial{#1}\over\partial{#2}}}

\newcommand{\pd}{\partial}

\newcommand\bfgrad{{\bs \nabla}}
\renewcommand{\v}[1]{{\bf #1}}

\newcommand{\cD}{ {\cal D} }

\newcommand{\cF}{ {\cal F} }

\newcommand{\cH}{ {\cal H} }
\newcommand{\cK}{ {\cal K} }

\newcommand{\cP}{ {\cal P} }

\newcommand{\cS}{ {\cal S} }

\newcommand{\gd}{ {\delta} }
\newcommand{\gD}{ {\Delta} }

\newcommand{\gq}{ {\theta} }

\newcommand{\gl}{ {\lambda} }
\newcommand{\gr}{\rho}

\begin{document}

 \bibliographystyle{unsrt}

%\tightenlines

\title{Numerical studies of entangled PPT states in composite quantum systems}

\author{Jon Magne Leinaas$^a$, Jan Myrheim$^b$ and Per {\O}yvind Sollid$^a$\\ 
${(a)}$ Department of Physics, University of Oslo,\\ P.O.
Box 1048 Blindern, NO-0316 Oslo, Norway\\
${(b)}$ Department of Physics, The Norwegian University of Science and Technology,\\ NO-7491 Trondheim, Norway}

\date{9 February, 2010}

\maketitle

\begin{abstract} 
We report here on the results of numerical searches for PPT states with specified ranks for density matrices and their partial transpose. The study includes several bipartite quantum systems of low dimensions. For a series of ranks extremal PPT states are found. The results are listed in tables and charted in diagrams. Comparison of the results for systems of different dimensions reveal several regularities. We discuss lower and upper bounds on the ranks of extremal PPT states.
\end{abstract}

%\pacs{}
%\maketitle

%%%%%%%%%%
\section{Introduction}

In recent years the study of entanglement in composite quantum systems has taken several different directions. One direction is the study of entanglement from a geometrical point of view \cite{Kus01,Verstraete02,Pittenger03,LeinaasMyrheim06,Bengtsson06}. This has led to questions concerning the relations between different convex sets of Hermitian matrices, where the full set of density matrices is one of them. The main motivation for the interest in these convex sets is the information they give about the general question of how to identify entanglement in a composite system. Unless the system is in a {\em pure} quantum state, the knowledge of the corresponding density matrix does not readily disclose the state as being entangled or non-entangled. And the complexity of the corresponding problem, known as the separability problem, increases rapidly with the dimensionality of the quantum system \cite{Gurvits03}.

The density operators are the positive, normalized, semidefinite Hermitian operators that act on the Hilbert space of the quantum system, and in the following we shall use the notation $\cD$ for this set. Another convex set is the set of {\em non-entangled} states, usually referred to as separable states, and we use $\cS$ as notation for this subset of $\cD$. Since the set of entangled states is the complement to the set of separable states, within the full set of density matrices, the question of identifying the entangled states can be reformulated as the question of finding the boundaries of the convex set of separable states $\cS$.

For a bipartite system there is furthermore a convex subset of the density matrices, here referred to as $\cP$, which is closely related to the set of separable matrices. This is the subset of density matrices that remain positive semidefinite under the operation of partial transposition of the matrix with respect to one of the two subsystems of the composite system. For short these states are called PPT states. A necessary condition for separability of a density matrix is that it remains positive under partial transposition, and thus the set of separable states is included in the set of PPT states,  $\cS\subset\cP$ \cite {Peres96}. For bipartite systems of dimensions 2x2 and 2x3 the two sets are in fact identical \cite{Horodecki96}, but in higher dimensions the separable states form a proper subset of the set of PPT states. However, numerical studies have shown that for systems of low dimensions, like the 3x3 system, the set $\cP$ is only slightly larger than $\cS$ \cite{LeinaasMyrheim06,LeinaasMyrheim07}.

The necessary condition that the separable density matrices remain positive under partial transposition is important, since this condition is easy to check. It effectively reduces the separability problem to a question of identifying the PPT states that are entangled, \ie that do not belong to $\cS$. These states are also interesting for 
%%%JML
a separate reason,
%%%JML
 since they are known to carry {\em bound} entanglement, which means that the entanglement is not available through entanglement distillation, a process where entanglement of mixed states is transferred to a set of pure quantum states \cite{Horodecki98}.  

All convex sets are in principle determined by their extreme points, which for $\cD$ are the pure quantum states and for $\cS$ are the pure product states. The set $\cP$ also has the pure product states as extreme points, but in addition there are other extreme points that are not fully known. In a previous publication \cite{LeinaasMyrheim07} we have found  a criterion for  identifying extremal PPT states, and we have described an algorithm to systematically search for such states. By use of the method, a list of extremal PPT states were found and presented for a series of low-dimensional systems. 

In the present paper we follow up the study of entangled PPT states in low-dimensional systems by use of numerical methods. The method described here is different from that of the publication \cite{LeinaasMyrheim07} in the sense that it does not make use of direct searches for extremal PPT states, but rather of searches for PPT states with specified ranks for the density matrix and for its partial transpose. 
By systematically searching through matrices of different ranks we have obtained in this way, for a series of low-dimensional systems, a list of low rank PPT states, many of which are identified as extremal and others as non-extremal entangled PPT states. These results supplement those of \cite{LeinaasMyrheim07}, where only a limited set of different types of extremal states were identified. 

The results we have obtained for different low-dimensional systems show certain regularities. In particular we find extremal PPT states for essentially all ranks of the density operator and its partial transpose, when these  lie between an upper and a lower limit. We suggest general expressions for these limits and relate them to generic properties of the image and kernel of the density matrices. A special focus is on the properties of the extremal PPT states of {\em lowest} rank. In a separate publication \cite{LeinaasMyrheim09} we follow up these results by a specific study of the lowest rank extremal PPT states of the 3x3 system (which are also identical to the lowest rang {\em entangled} states in higher dimensions \cite{Horodecki00}) . As discussed there, these  states can be classified by a small number of parameters, and an interesting question is whether a similar classification can be given for other extremal PPT states .

%%%%%%%%%%
\section{The method}

We consider a bipartite quantum system with Hilbert space $\cH=\cH_A\otimes\cH_B$, where $A$ and $B$ label the two subsystems, and $\cH_A$ and $\cH_B$ are of dimensions $N_A$ and $N_B$ respectively. 
The density operators $\gr$ satisfy the normalization condition $\Tr \gr=1$, but often it is convenient to give up this condition and rather consider all density operators that differ by a normalization factor to be equivalent. The set of normalized density operators we refer to as $\cD$, while $\cK(\cD)$ is the positive cone of non-normalized density operators. A similar notation is used for other subsets of Hermitian operators.
Partial transposition is the operation on density operators
that corresponds to transposition of indices of one of the
subsystems, $\gr^B _{ij}\to\gr^B_{ji}$ (here chosen as subsystem
$B$). For density operators of the full, composite system, we refer to
this operation as $\gr\to\gr^P$, and it maps the convex set $\cD$ into another convex set denoted $\cD^P$. The operation will depend on the
choice of subsystem ($A$ or $B$) and on the choice of basis in the
corresponding Hilbert space. However, the distinction between these
different choices is of no importance for our discussion. We note in
particular that the mapping of sets $\cD\to\cD^P$ is independent of
the choice.
The set of PPT states is defined as the section $\cP=\cD\cap\cD^P$, which means that it consists of the positive semidefinite operators that remain positive semidefinite under partial transposition. 

\subsection{Searching for density operators of given ranks}

The Hermitian matrices define a real vector space, and it is convenient to introduce  a complete set of matrices that are orthonormal with respect to the trace norm
\be{norm}
\Tr(M_iM_j)=\gd_{ij}
\ee
A general Hermitian matrix M is then described as a vector $\v x$ with real components
\be{comp}
x_i=\Tr(M M_i)
\ee
For a composite system of Hilbert space dimension $N=N_AN_B$ the vector space of Hermitian matrices is of dimension $N^2=N_A^2N_B^2$.

The algorithm we use to find PPT states $\gr$ with specified ranks $(m,n)$ for $\gr$ and $\gr^P$ is the following. 
We expand $\gr$ as
\be{JM1}
\gr=\gr(\v x)=\sum_ix_iM_i
\ee
The eigenvalues of $\gr$ we write as  $\gl_i=\gl_i(\v x)$, with $\gl_i^P=\gl_i^P(\v x)$ as the eigenvalues of the partially transposed matrix $\gr^P$.  The eigenvalues of each matrix are listed in
decreasing order, and for the density matrix that we are searching for, a certain number of the eigenvalues should vanish. Thus we want to have $\gl_k=0$ for $ k=m+1,...,N$ and $\gl_k^P=0$ for $k=n+1,...,N$. The eigenvalues that should vanish we treat as components of a new vector $\bs \mu$, so that
\be{muvec}
\bs\mu=[\gl_{m+1},\gl_{m+2},...,\gl_N,\gl^P_{n+1},\gl^P_{n+2},...,\gl_N^P]
\ee
and the problem is then to find the point $\v x$ which solves the equation  $\bs\mu(\v x)=0$. 

 We choose a starting point $\v x$ such that $\gr=\gr(\v x)$ as well as
$\gr^P$ are positive semidefinite matrices (it is not strictly
necessary that the positivity conditions hold to begin with, since
they will automatically hold for the solution we obtain in the end).
If the equation $\bs\mu(\v x)=0$ is not already solved, we search for
a better approximate solution $\v x'=\v x+\gD \v x$.  The linear
approximation to the Taylor expansion gives an equation
for $\gD \v x$,
\be{lin}
\bs\mu(\v x)+(\gD\v x\cdot\bfgrad)\bs\mu(\v x)=0
\ee
In matrix form the equation can be written as
\be{mat}
\bs B\,\gD\v x=-\bs\mu
\ee
with 
$B_{ij}=\pd\mu_i /\pd x_j$. It implies another equation,
\be{mat2}
\bs A\,\gD\v x=\v b
\ee
where $\bs A=\bs B^T\bs B$ is a positive, real symmetric matrix, and
where $\v b=-\bs B^T\bs\mu$.  We use the conjugate gradient method to
solve the last equation for $\gD\v x$.  The conjugate gradient method
is useful because it works even if $\bs A$ is singular.  Next, we
replace $\v x$ by $\v x'=\v x+\gD\v x$, and iterate in order to get
successively better approximate solutions.

The matrix $\bs B$ is computed in each iteration by the formula
\be{iden}
\pdd{\gl_k}{x_j}=\psi_k^\dag\,\pdd{\gr}{x_j}\,\psi_k
=\psi_k^{\dag}\,M_j\,\psi_k\,,\quad k =m+1,...,N_A
\ee
where $\psi_k$ is the eigenvector of $\gr$ with eigenvalue $\gl_k$.
A similar formula is used for the derivatives of $\gl_k^P$. Since these
formulae are valid in first order {\em non-degenerate} perturbation
theory, this raises a question concerning convergence of the method at a point of degeneracy. However, in practice we find that the method works well when the dimension of the system is not too large.

Note that, by the way the method works, all the states we find are PPT. That is the case since in the iterative search for a state with a certain number of vanishing eigenvalues for $\gr$ and $\gr^P$, the eigenvalues are always ordered in such a way that {\em the lowest}
eigenvalues are forced to be zero.  This means that both the density
matrix and its partial transpose will be positive
semidefinite.

We have applied the method to a series of low-dimensional systems, and the results are listed in the tables in Appendix B and in the diagrams in the next section. The convergence of the method slows down with increase of the Hilbert space dimension $N$, and in the form we have implemented the algorithm, the practical limit of the dimension is $N\lesssim 20$. For the systems we have studied, the method has been used repeatedly with different starting points for each choice of ranks $(m,n)$. In most cases the iteration converges, but in some cases that does not happen and these iterations are then simply aborted. For some values of $m$ and $n$ we do not find any density matrix with the given ranks, and the method will then in most cases converge to a density matrix with lower rank for either $\gr$ or $\gr^P$, or both. We have not imposed any restriction to avoid that, so in practice the method searches for density matrices of ranks equal to or {\em lower} than the specified values $(m,n)$.

%%%%%%%%%%
\subsection{Determining the dimension of the face of $\cK(\cP$)}

For each density matrix $\gr$ found in the searches, we have evaluated and listed certain properties. Among these are the {\em local ranks} $(r_A,r_B)$ of the density matrices, defined with respect to the subsystems $A$ and $B$. These are the ranks of the reduced density matrices $\gr_A$ and $\gr_B$. The most interesting cases are those where the density matrices have full local ranks $(r_A,r_B)=(N_A,N_B)$. If that is not the case the density matrix can be viewed as belonging to a composite system of lower dimension, which is embedded in the higher dimensional system. 

We have also evaluated and listed the dimension of the {\em face} $\cF$ of the convex cone $\cK(\cP)$ to which $\gr$ belongs. 
The extreme points of $\cP$ correspond to one-dimensional, positive {\em rays} in  $\cK(\cP)$, and are consequently characterized by $\dim\,\cF=1$. Therefore the extremal states can be identified in the tables as the density matrices with this minimal value for the dimension of the face.

The method we use to evaluate $\dim\cF$ has earlier been described in \cite{LeinaasMyrheim07}. It is based on the fact that the face $\cF$
to which $\gr$ belongs can be viewed as the section between a face (or all) of
$\cK(\cD)$  and a face (or all) of $\cK(\cD^P)$. This means that $\gr=\gr(\v x)$ satisfies two
equations,
\be{twoeq}
\bs P\v x=\v x\,,\quad \bar{\bs Q}\v x=\v x
\ee
with $\bs P$ as the orthogonal projection on the subspace in the
vector space of Hermitian matrices defined by the face of $\cK(\cD)$,
and $\bar{\bs Q}$ as the projection on the subspace defined by the
face of $\cK(\cD^P)$. Note that $\bs P$ and $\bar{\bs Q}$ are real and
symmetric $N^2\times N^2$ matrices. The method for computing these
matrices for a given density operator $\gr$ is described in
\cite{LeinaasMyrheim07}.

As follows from the above equations, the section between the subspaces defined by $\bs P$ and $\bar{\bs Q}$ is spanned by the eigenvectors of the composite, real symmetric matrix $\bs P\bar{\bs Q}\bs P$ (or alternatively $\bar{\bs Q}\bs P\bar{\bs Q}$), with {\em eigenvalues equal to $1$}. This implies that $\dim\,\cF$ can be determined simply by diagonalizing the composite matrix and counting the number of such eigenvalues.

As a consequence of the way $\cF$ is constructed, there is a geometrical constraint on the possible values of $\dim\,\cF$ for given ranks $(m,n)$ of $\gr$ and $\gr^P$  \cite{LeinaasMyrheim07}. Thus, the faces of $\cK(\cD)$ and $\cK(\cD^P)$ are of dimensions $m^2$ and $n^2$ respectively, and therefore the face of $\cK(\cD)$ is specified by $N^2-m^2$ linear constraints and similarly the face of $\cK(\cD^P)$ is specified by $N^2-n^2$ linear constraints. If these sets of constraints are independent they determine the dimension of the section $\cF$ as ${\rm dim}\,\cF= m^2+n^2-N^2$. However, if the two sets of constraints are not fully independent, the dimension of the section is larger. Therefore the following inequality is generally valid,
\be{bound}
{\rm dim}\,\cF\geq m^2+n^2-N^2
\ee
For extremal PPT states, with $\dim\,\cF=1$, this gives an upper bound to the ranks of $\gr$ and $\gr^P$,
\be{ineq}
m^2+n^2\leq N^2+1\quad\quad {\rm (extremity)}
\ee

\subsection{Counting the number of product vectors}

For each density matrix $\gr$, we have examined the image ($\Im\gr$) and kernel ($\Ker\gr$) for the presence of product vectors, and the numbers of such vectors are listed. The method we use to search for product vectors in a given subspace of $\cH$ is described in some detail in Appendix A. To briefly outline the approach, let us assume that we search for product vectors in $\Im\gr$, with $P$ as the orthogonal projection on this subspace. The vectors should satisfy the condition  
\be{pv}
(\iden - P)(\phi\otimes\chi)=0
\ee
and this can be re-expressed as a minimization problem, which we in the Appendix refer to as a {\em double eigenvalue problem}. The solutions of \pref{pv}  are  the minima of the function 
\be{func}
f=(\phi^\dag\otimes\chi^\dag)(\iden-P)(\phi\otimes\chi)
\ee
with
$f=0$,
and such minima can be found by the iterative approach described in
the Appendix. By varying the starting point of the iteration,
different minima can be identified, and by systematically searching
for minima of $f$ and of
\be{funcI}
1-f=(\phi\otimes\chi)^\dag P(\phi\otimes\chi)
\ee
we have reproduced and counted the product vectors
in $\Im\gr$ and $\Ker\gr$ for
every density matrix
found in the numerical searches. In the tables the number of product
vectors and the number of {\em linearly independent} product vectors
are listed for $\Im\gr$ and $\Ker\gr$.

The same type of iterative method for minimization over product
vectors has been applied in the separability test that is described in
the next section.

%%%%%%%%%%
\section{Discussion of the results}

\subsection{Diagrammatic representation}

The results of the searches are tabulated in Appendix B, and are also included in condensed form in the text below, as the diagrams in Figs.~\ref{Diagram}-\ref{Diagram4}.  
The  variables of the two axes of the diagrams are the ranks $m$ of $\gr$ and $n$ of $\gr^P$. The small open and filled circles in the diagrams indicate the ranks for states that are found in the numerical searches. As discussed above, all the states produced in this way are PPT. The filled (red) circles represent {\em extremal} PPT states with full local ranks, while the open circles represent non-extremal PPT states, either separable or entangled.
Note that all diagrams are symmetric under the interchange $m\leftrightarrow n$, since in the search for PPT states there is no intrinsic difference between $\gr$ and $\gr^P$.

The states we find by use of our method we refer to as {\em typical} for the chosen ranks $(m,n)$. For some ranks there will exist states with untypical characteristics, which we do not find in the searches,  presumably due to low dimensionality of the set of such states.  We note in particular, when we compare the diagrams of the 3x3 and 4x4 systems, that most of the states of the 3x3 system are not seen in the diagram of the 4x4 system. However, we know that all the states of the lower-dimensional system should be present,  in the form of states with less than full local ranks. 

For most choices of ranks $(m,n)$ the density matrices that we find in repeated searches with different initial conditions, are all found to have identical characteristics in the form of the parameters listed in the tables. There is only a small number of exceptions where density matrices with different characteristics but equal ranks $(m,n)$ have been found. These are all listed in the tables. 

\subsection{Different groups of PPT states}
\begin{figure}[h]
\begin{center}
\includegraphics[width=12cm]{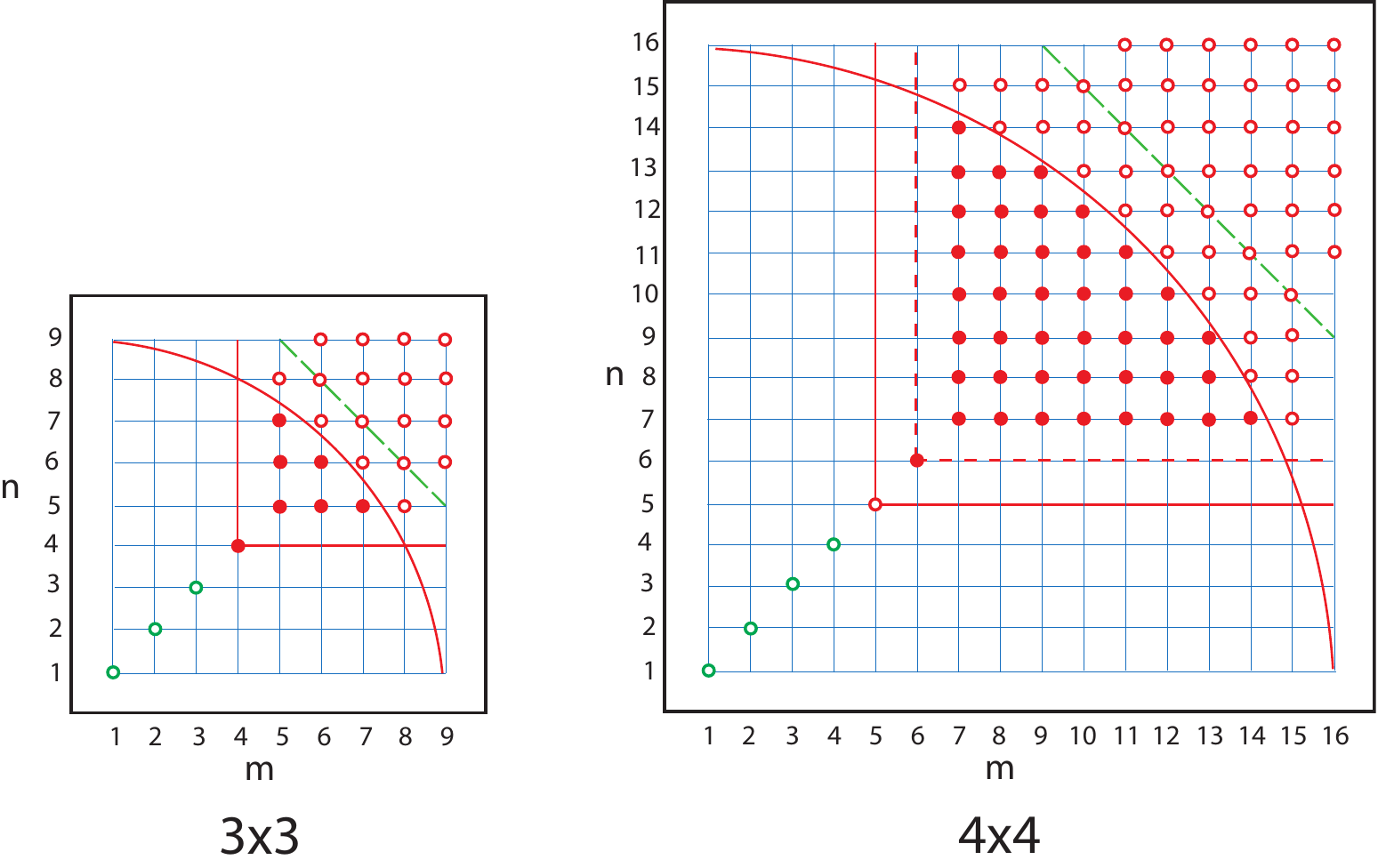}
\end{center}
\caption{\small   Diagrams for the ranks of PPT states in the composite
  systems of dimensions 3x3 and 4x4. The coordinates
  $m$ and $n$ on the axes are the ranks of the
  density matrix $\gr$ and its
  partial transpose $\gr^P$. States found by use of
  the numerical method discussed in the text are indicated by small
  circles or dots (filled circles).  The green circles
  in the lower left corner
  indicate ranks for a special set of separable states, the red dots
  correspond to extremal PPT states
  and the red circles
  in the upper right corner
  to non-extremal PPT states.
  Ranks $(5,5)$ in the 4x4 system is an exception, there we find both
  separable states and a special type of entangled, non-extremal PPT states.
  The unbroken,
  horizontal and vertical straight
  red lines
  show the lower bound for entangled PPT states with full local ranks,
  the similar dashed  red lines
  show our conjectured lower bound for extremal
  PPT states with full local ranks, and the red circular arcs
  show the upper bound for extremal PPT states. The dashed green
  $45^{\circ}$ straight lines
  show the upper bound for the application of the separability
  criterion described in the text. For all ranks indicated by
  red circles up to and on this line (and above the circular arc),
  the corresponding states are always found to be entangled PPT states.  \label{Diagram}}
\end{figure}

\begin{figure}[h]
\begin{center}
\includegraphics[width=13cm]{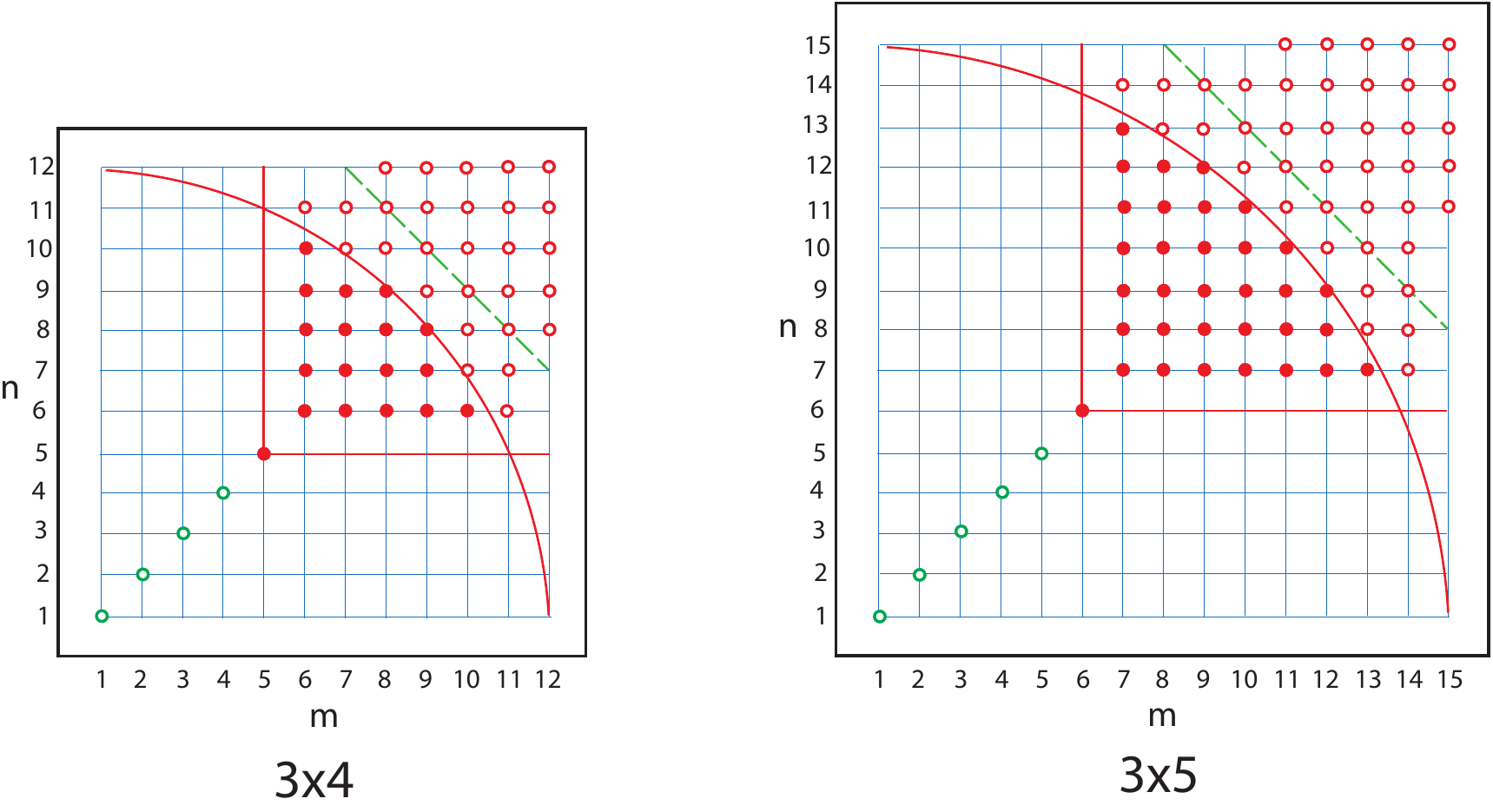}
\end{center}
\caption{\small Diagrams for the ranks of PPT states in the composite systems of dimensions 3x4 and 3x5.  States found by use of the numerical method discussed in the text are indicated by the small circles or dots. Also here entangled PPT states are found for all ranks indicated by red circles up to and on the dashed green line. For further explanation of the diagrams see Fig.~\ref{Diagram}.\label{Diagram2}}
\end{figure}

\begin{figure}[h]
\begin{center}
\includegraphics[width=10cm]{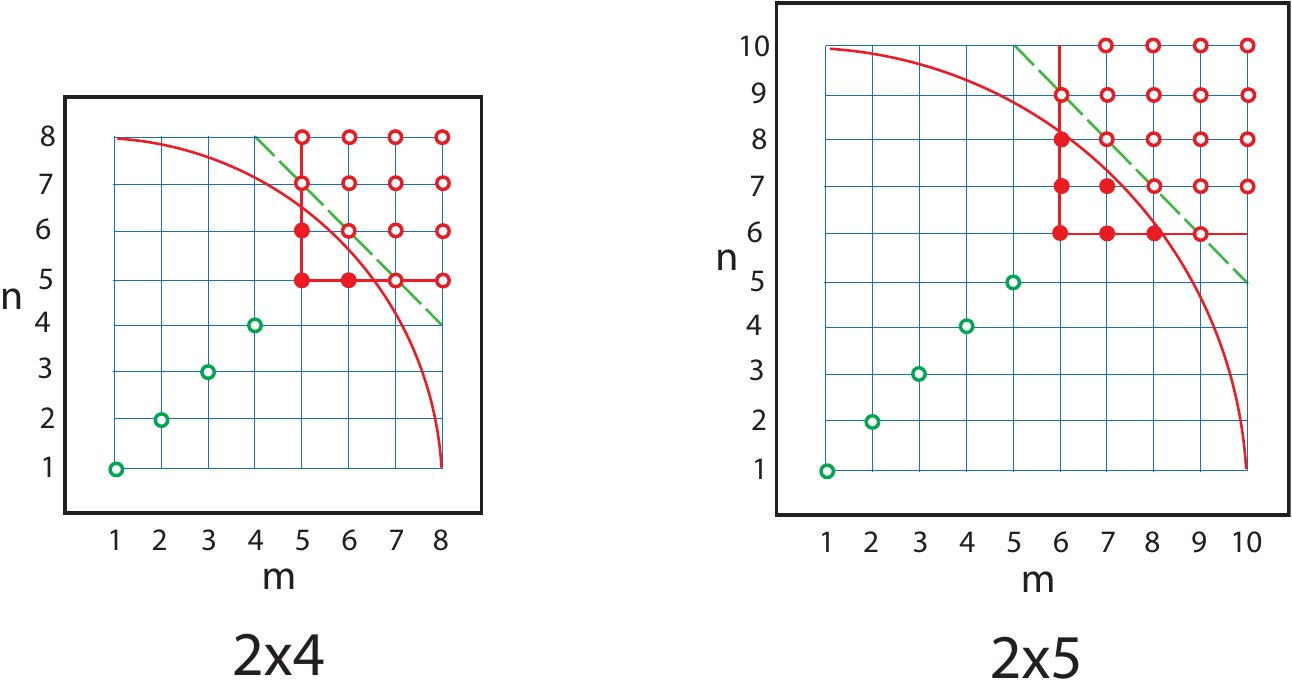}
\end{center}
\caption{\small Diagrams for the ranks of PPT states in the composite systems of dimensions 2x4 and 2x5.  For explanations of the diagrams see Figs.~\ref{Diagram} and \ref{Diagram2}.\label{Diagram3}}
\end{figure}

\begin{figure}[h]
\begin{center}
\includegraphics[width=9cm]{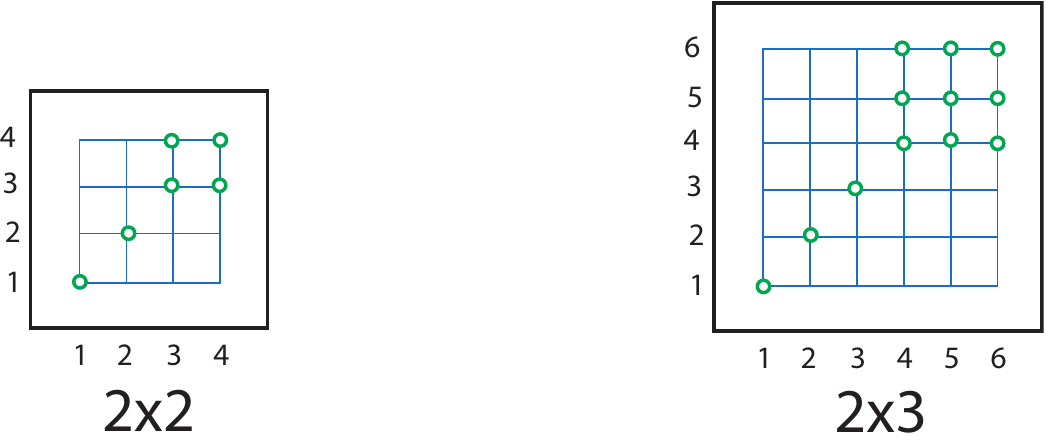}
\end{center}
\caption{\small Diagrams for the ranks of PPT states in the composite systems of dimensions 2x2 and 2x3.  For these systems all PPT states are separable, but the distinction between the special set of states with low, symmetric ranks $m=n$ and the set of states with higher, asymmetric ranks is similar to what is found in the higher-dimensional systems.\label{Diagram4}}
\end{figure}

There is a clear similarity between the diagrams of
Figs.~\ref{Diagram}-\ref{Diagram4}, with the states for each diagram
being separated into groups with different characteristics.  One group
consists of the low-rank states with
$m=n\leq{\rm max}\{N_A,N_B\}$. These low-rank states, which are represented by the series of green circles in the diagrams, are all separable, with equal ranks for $\gr$ and $\gr^P$. As shown in the tables, all the corresponding density matrices $\gr$ contain a number of product vectors in their image which is equal to their rank $m$. In fact, it is obvious that one can construct such states by taking randomly a small number of pure product states and forming a convex combination of these. 
Such a sequence of separable states can be defined in all dimensions, and they are not of further interest for our discussion. 
%%%%%%%%%%
%%%%%%%%%

The remaining states, with ranks $m> {\rm max}\{N_A, N_B\}$ and $n> {\rm max}\{N_A, N_B\}$, are restricted to the upper right corners of the diagrams. For these values of the ranks, all states produced in our searches, with one exception, have full local ranks. (The exception is the case $(m,n)=(5,5)$ of the 4x4 system, where in addition to full rank states also states with less than full rank are found.) The systems of dimension 2x2 and 2x3, represented by the diagrams in Fig.~\ref{Diagram4}, are special, since for these systems all PPT states are separable. However, for the higher-dimensional systems we note that the above restriction on the ranks of $\gr$ and $\gr^P$ coincides with a lower bound, found by P. Horodecki \etal \cite{Horodecki00}, on the rank of {\em entangled} PPT states with full local ranks.  (In the following we shall refer to this as the HLVC bound.) In fact, with the exception of some of the $(5,5)$ states of  the 4x4 system, we find for all ranks above and sufficiently close to this lower bound that the states are not only entangled but extremal PPT states.

As displayed by the diagrams, there are a few intermediate values of $m$, above the HLVC bound, where only symmetric matrices, $m=n$ are found. We shall discuss these states separately in Sect.~3.4, and focus now on the states with ranks $m\geq N_A+N_B-1$ and $n\geq N_A+N_B-1$.  For almost all ranks that satisfy these inequalities we find PPT states.  The only exceptions are some of the cases with the largest asymmetry between the values of $m$ and $n$. In fact we cannot rule out that we miss these states as a consequence of the method we use. We find that with large asymmetry in $m$ and $n$ the numerical method seems preferably to pick up matrices with lower and more symmetric ranks. Therefore a much larger number of searches have to be performed in order to find density matrices when the ranks are highly asymmetric.

With exception for the systems 2x2 and 2x3, all the states we find with ranks between the lower bounds $m\geq N_A+ N_B-1$ and $n\geq N_A+ N_B-1$ and the upper bound $m^2+n^2\leq N^2+1$ are {\em extremal} (and hence entangled) PPT states. The upper bound is the constraint \pref{ineq} on extremity that has already been discussed, and in the diagrams this bound is displayed as the red circular arc. For ranks above this bound, we have used a separability test \cite{Horodecki00} on the states, and this test shows that for all ranks that satisfy the inequality $m+n\leq 2N-N_A-N_B+2$, we find entangled PPT states. Beyond this limit the test is not applicable.

%%%%%%%%%%
%%%%%%%%%

%%%%%%%%%%
\subsection{Dimensions of faces and numbers of product vectors}
%%%%%%%%%%
The structure of the diagrams discussed above can to some extent be related to simple regularities of the parameters in the tables. We focus first on the list of the values of $\dim \cF$. This number is constrained by the geometric bound \pref{bound}, $\dim\,\cF\geq m^2+n^2-N^2$. We note that for all states where the ranks are sufficiently large to give a positive number for this lower bound, the constraint is satisfied with equality.  This means that the linear constraint equations \pref{twoeq} that define $\cF$ as a section between faces of $\cK(\cD)$ and $\cK(\cD^P)$
are all independent. 

For all ranks $m,n\geq N_A+N_B-1$ which give negative values for the bound, we find $\dim \cF=1$. This is the minimum value consistent with the fact that the density operator $\gr$ is located on both faces.  Other "accidental" relations between the two intersecting faces, therefore, seem not to be present. The fact that $\dim\,\cF$ takes the minimal value consistent with this condition gives an explanation for why all the states with ranks between a lower and an upper bound are found to be extremal PPT states.
For sufficiently low ranks, $m,n< N_A+N_B-2$, we find states where $\dim\,\cF$ does not take this minimum value. These states we therefore see as corresponding to more special constructions.

The numbers of product vectors we find in $\Im\gr$ and $\Ker\gr$ also show a simple regularity.  For a long sequence of high ranks $m$ all states have no product vector in $\Ker\gr$ and a complete basis (in fact an overcomplete set) of product vectors in $\Im\gr$. When lowering the rank there are in some of the lists a small set of intermediate ranks where there is no product vector in neither $\Im\gr$ nor $\Ker\gr$, and below this there is a single extremal state at $m=n=N_A+N_B-2$ (not present in the $2\times N_B$ systems) with a complete set of product vectors in $\Ker\gr$ and no product vector in $\Im\gr$. For even lower ranks the states are also here exceptional and do not fit into this picture.

 %%%%%%%%%%
%%%%%%%%%
When we exclude these lowest-rank states, the numbers we find in the lists are in fact identical to the numbers of product vectors expected for generic subspaces of the given dimension. To show this we consider the conditions for a product state $\phi\otimes\chi$ to be present in a randomly chosen subspace of dimension $d$ in the Hilbert space $\cH=\cH_A\otimes\cH_B$. The product state will satisfy a set of constraint equations of the form
\be{cons}
\psi_k^\dag(\phi\otimes\chi) = 0\,,\quad k=1,2,...,N-d
\ee
where $\psi_k$ is a linearly independent set of states that are all orthogonal to the chosen $d$-dimensional space. The solutions to the equation will generally depend on a number of continuous parameters that can be determined by parameter counting. Thus, the product state will be specified by $N_A+N_B-2$ complex parameters (when the complex normalization factors are not included), and this number is reduced by the $N-d$ complex constraint equations to give for the solution a remaining set of $p$ complex parameters, with
\be{para}
p=N_A+N_B-2-N+d
\ee 
We may then distinguish between three cases,
\vskip1.5mm

\noi
1) $p>0$, which means $d>N-N_A-N_B+2$: The set of equations \pref{cons} is underdetermined, and there is an infinite set  of product vectors in the $d$-dimensional subspace, described by $p$ complex free parameters.\vskip1mm
\noi
2) $p=0$, which means $d=N-N_A-N_B+2$: The number of equations matches the number of parameters to give a finite set of solutions.
\vskip1mm
\noi
3) $p<0$, which means $d<N-N_A-N_B+2$: The set of equations is overdetermined and there is in the generic case no solution. There may however be solutions for specially selected subspaces.
\vskip1.5mm

For the case 2) the number of product vectors is given by the expression
\be{nosol}
n_{ps}= \pmatrix{N_A+N_B-2\cr N_A-1}=\frac{(N_A+N_B-2)!}{(N_A-1)!(N_B-1)!}
\ee
The problem of finding this number of product vectors can be related to the problem in algebraic geometry of finding the degree of the variety defined by the Segre embedding between projective spaces $P^{N_A-1}\times P^{N_B-1}\to P^{N_AN_B-1}$ \cite{Hartshorne06}. The number given above is identical to this degree.

It is straight forward to check that all the numbers in the tables are consistent with these results, except for the special low-rank states.  Apart from these states, the PPT states we find in the numerical searches are therefore also in this respect typical states for the given ranks $(m,n)$.

 %%%%%%%%%%
%%%%%%%%%

\subsection{The lowest-rank extremal and entangled PPT states}

For the systems with $N_A>2$ and $N_B>2$ the lowest-rank extremal PPT states with full local ranks seem to have a special status. These states have symmetric ranks $m=n$, and for these values of $m$ (or $n$) we find no  states with asymmetric ranks. They are also special in the sense that they are the only ones with no product vector in $\Im\gr$ and a finite, complete set of product vectors in $\Ker\gr$. 

For the 3x3 system these are rank $(4,4)$ states, and the presence of
product vectors in $\Ker\gr$ but not in $\Im\gr$ indicates that they are related to a special construction of low rank entangled PPT states for this system, with the use of {\em unextendible product bases} (UPB) \cite{Bennett99}. The UPB is a set of orthogonal product vectors that spans a subspace ($\Ker\gr$), and that cannot be extended with additional orthogonal product vectors (in $\Im\gr$). The $(4,4)$ states that we find by our method do not directly exemplify this construction, since the product vectors in $\Ker\gr$ are non-orthogonal, but there is a general connection to the UPB construction that we discuss in detail in a separate publication \cite{LeinaasMyrheim09}.

We find similar types of extremal PPT states in all the systems we have studied with dimensions of the subsystems larger than 2. This we take as an indication for the presence of such lowest-rank extremal states in all higher-dimensional systems. When we combine the assumption of a finite, complete set of product vectors in $\Ker\gr$ and no product vector in $\Im\gr$ with the expectations for the number of product vectors in generic subspaces of a given dimension, we find that these states should generally have  ranks equal to $m=n= N_A+N_B-2$.
We will phrase the assumption about these states in the form of the following conjecture:
\vskip1mm
{\em The lowest-rank extremal PPT state with full local ranks in an $N_A \times N_B$ system, with $N_A$ and $N_B$ larger than $2$, is characterized by symmetric ranks m=n for the density matrix and its partial transpose, with value $m=n=N_A+N_B-2$.}
\vskip1mm

We write this as the following inequality for the rank of {\em extremal} PPT states with full local ranks, 
\be{con}
m,n\geq N_A+N_B-2\quad\quad ({\rm conjecture})
\ee
and compare it to the lower bound for {\em entangled} PPT states with full local ranks,
\be{Horo}
m,n\geq {\rm max}\{N_A,N_B\}+1\quad\quad ({\rm HLVC})
\ee
For systems of dimension $2\times N_B$ we note that the HLVC bound lies above the lower bound for extremity,  and therefore the bound \pref{con} cannot be saturated. That is in accordance with the lack of the special type of lowest-rank extremal PPT state for these systems. For systems of dimension $3\times N_B$ (with $N_B\geq 3$) the two bounds coincide, and the lowest rank extremal states that we have found for these systems are indeed also the lowest rank entangled states. For systems of dimension $4\times N_B$ (with $N_B\geq 4$) there is a difference of $1$ between the two lower bounds with the HLVC bound being the lowest. We have examined only one example of such systems, namely the 4x4 system. For this system we do find entangled PPT states with lower rank than the lowest-rank extremal PPT state. For general systems $N_A\times N_B$, with $N_A\leq N_B$, the difference between the two bounds is according to our conjecture equal to $N_A-3$, and therefore increases linearly with the lowest dimension of the two subsystems. 

The entangled PPT states with lower ranks than the lowest rank
extremal PPT states with full local ranks will include only extremal
PPT states with less than full local ranks, when written as convex
combinations of extremal states. We have found examples of such states
with ranks $(5,5)$ in the 4x4 system. In the table such a state $\gr$
appears with $\dim\cF=2$, which means that it is a convex combination
of two extremal PPT states,
\be{concom}
\gr=(1-x)\gr_e+x\gr_p
\ee
with $0<x<1$.  One of these is a pure product state $\gr_p=ww^{\dag}$,
where $w=u\otimes v$ is the single product vector found in
$\Im\gr$. The other one is a rank $(4,4)$ state $\gr_e$, which we
identify as an extremal PPT state of a 3x3 subsystem. 
We find this state by subtracting $\gr_p$ with the value of the coefficient $x$ determined as discussed in \cite{Horodecki00}. 
To summarize, the entangled rank $(5,5)$
PPT state $\gr$ has full local ranks and saturates the HLVC bound, but
is a convex combination of two extremal PPT states that both have less
than full local ranks.

A closer look at the above decomposition of the $(5,5)$ state $\gr$,
motivates the following general construction of states that saturate the HLVC bound
in successively higher dimensions.  Assume that
\be{decompHAHB}
{\cal H}_A={\cal U}_1\oplus{\cal U}_2\;,\qquad
{\cal H}_B={\cal V}_1\oplus{\cal V}_2
\ee
where $\dim{\cal U}_1=\dim{\cal V}_1=3$ and
$\dim{\cal U}_2=\dim{\cal V}_2=1$.  We assume that ${\cal U}_1$ and
${\cal U}_2$ are complementary but not necessarily orthogonal
subspaces of ${\cal H}_A$, and that ${\cal V}_1$ and ${\cal V}_2$ are
complementary but not necessarily orthogonal in ${\cal H}_B$.

Let $\gr_e$ be a rank $(4,4)$ extremal PPT state on the $3\times 3$
dimensional subspace ${\cal U}_1\otimes{\cal V}_1\subset{\cal H}$.  A
property of $\gr_e$ is that there are no product vectors in
$\Im\gr_e\subset{\cal U}_1\otimes{\cal V}_1$.  Let
$\gr_p$ be a pure product state, $\gr_p=ww^{\dag}$ with
$w=u\otimes v$, $u\in{\cal U}_2$ and $v\in{\cal V}_2$.  Then define
$\gr=(1-x)\gr_e+x\gr_p$ with $0<x<1$.  The image of $\gr$ is
\be{Imrhocc}
\Im\gr=\Im\gr_e\oplus\Im\gr_p
=\Im\gr_e\oplus({\cal U}_2\otimes{\cal V}_2)
\subset({\cal U}_1\otimes{\cal V}_1)\oplus({\cal U}_2\otimes{\cal V}_2)
\ee
Since ${\cal U}_2\otimes{\cal V}_2$ is one dimensional, it follows
that the rank of $\gr$ is one higher than the rank of $\gr_e$.
Similarly, the rank of $\gr^P$ is one higher than the rank of
$\gr_e^P$.

The only product vector in $\Im\gr$ is now $w=u\otimes v$.  To see
this, consider the most general product vector
\be{prodvec11}
w'=(u_1+u_2)\otimes(v_1+v_2)
=(u_1\otimes v_1)+(u_1\otimes v_2)+(u_2\otimes v_1)+(u_2\otimes v_2)
\ee
with $u_i\in{\cal U}_i$ and $v_j\in{\cal V}_j$.  It is a sum of four
vectors belonging to four complementary subspaces
${\cal U}_i\otimes{\cal V}_j\subset{\cal H}$ with $i,j=1,2$.  In order
to have $w'\in\Im\gr$ we must have
\be{pvcond}
w'\in({\cal U}_1\otimes{\cal V}_1)\oplus({\cal U}_2\otimes{\cal V}_2)
\ee
but this requires that $u_1\otimes v_2=u_2\otimes v_1=0$.   Since we want to have $w'\neq 0$, the only possibilities
left are $w'=u_1\otimes v_1$, which is not in $\Im\gr$, or
$w'=u_2\otimes v_2$, which is not new.

It should be clear from this analysis that the same construction can
be applied to higher dimensional systems.  For example, we may
reinterpret $\gr_e$ to be the $(5,5)$ state $\gr$ of the 4x4 system,
as constructed above.  We reinterpret $\gr_p$ to be a new pure product
state that increases the local ranks from 4 to 5, in a similar manner
as discussed, so that $\gr=(1-x)\gr_e+x\gr_p$ is a $(6,6)$ state of
the 5x5 system. The analysis of product vectors in $\Im\gr$ can be
done in the same way, with one exception. In this case there is
already one product vector in $\Im\gr_e$ and therefore there are two
possible solutions for the product vector $w'$, namely the newly added
vector $w=u\otimes v$ and the previous product vector in
$\Im\gr_e$. Since the number of product vectors is lower than
$\dim\Im\gr$ the state is entangled, and it saturates the HLVC bound.

The construction can be continued to arbitrarily high dimension, and
it will always create an entangled state that saturates the HLVC
bound. The state will be a convex combination of extremal PPT states
with less than full local ranks, but it will itself have full local
ranks. The construction is not restricted to symmetric cases,
$N_A=N_B$, since asymmetric systems can be reached by introducing
product vectors which increase the dimension of only $\cH_A$ or
$\cH_B$. In this way it is possible to saturate the HLVC bound in all
higher-dimensional systems.

%%%%%%%%%%
\subsection{Checking for entanglement at higher ranks}
All the states with ranks above the upper limit for extremity, \ie with
\be{ineq2}
m^2+n^2 > N^2+1
\ee
we find to have a complete set of product vectors in their image. This means that they satisfy the necessary condition for separability given by the range criterion, which however, does not exclude the possibility that they may be entangled. To get some further information about this we have made  use of a separability criterion introduced in \cite{Horodecki00}. This condition for separability can be seen as a strengthened version of the range criterion, and is based on a relation that, for separable states, exists between product vectors in $\Im\gr$ and in $\Im\gr^P$. We describe below the basis for this criterion and further describe the method we have applied for checking the criterion. 

Assume $\gr$ to be a separable density operator, which therefore can be written as a convex combination of product states,
\be{conexp}
\gr=\sum_k p_k \psi_k\psi_k^\dag
\ee
with $\psi_k=\phi_k\otimes\chi_k$. The partially transposed density operator can then be written as
\be{conexp2}
\gr^P=\sum_k p_k \tilde\psi_k\tilde\psi_k^\dag
\ee
with $\tilde\psi_k=\phi_k\otimes\chi^*_k$, where $\chi^*_k$ is the complex conjugate of $\chi_k$ with respect to the same basis in $\cH_B$ that is used for the partial transposition.  Therefore, corresponding to the set of product vectors $\{\psi_k\}$, which spans $\Im \gr$, there is a set of product vectors $\{\tilde\psi_k\}$, which we will refer to as the conjugate set, that spans $\Im \gr^P$. This implies that a necessary condition for separability of a density operator $\gr$ is that the number of pairs of product vectors in $\Im\gr$ and in $\Im \gr^P$ {\em that are conjugate}, is equal to or larger than the ranks of both $\gr$ and $\gr^P$. We write the condition as
\be{sepcond}
K\geq {\rm max}\{m, n\}
\ee
with $K$ as the number of conjugate pairs of product vectors in $\Im\gr$ and $\Im\gr^P$.

The above condition for separability is effectively restricted to cases where the ranks $m$ and $n$ of $\gr$ and $\gr^P$ are not too large. This follows since, if the dimensions of $\Im\gr$ and $\Im\gr^P$ are sufficiently large, the number of such pairs will necessarily be infinite. The condition for the number of pairs to be finite can be determined by essentially the same method of counting parameters as used to determine the typical number of product vectors in a Hilbert space of given dimension, as discussed in Sect.~3.3. Thus, a conjugate pair of product vectors, $(\psi,\tilde\psi)$, with $\psi=\phi\otimes\chi$ and $\tilde\psi=\phi\otimes\chi^*$, has to satisfy two sets of equations,
\be{orthocond}
\theta_i^\dag\psi=0\,,\quad \xi_i^\dag\tilde\psi=0
\ee
where $\{\gq_i\}$ is a basis of $\Ker\gr$ and $\{\xi_i\}$ is a basis of $\Ker\gr^P$. The number of equations that $\psi$ has to satisfy will be equal to, or will exceed the number of free parameters in this product state, provided the following condition is satisfied $\cite{Horodecki00}$
\be{critlim1}
\dim \Ker \gr+\dim\Ker \gr^P \geq \dim\cH_A+\dim\cH_B-2
\ee 
If the condition is satisfied with proper inequality there will typically be no solution and if it is satisfied with equality there will be a finite set of solutions. 
Expressed in terms of the ranks $m$ and $n$ and the Hilbert space dimensions $N$, $N_A$ and $N_B$ the inequality takes the form 
\be{critlim2}
m+n\leq 2N-N_A-N_B+2
\ee
and this sets the limit for applications of the criterion. In the diagrams this limit is indicated by the dashed green line.

For states that satisfy the above inequality, we have found a practical method to check the separability condition \pref{sepcond} by applying essentially the same double-eigenvalue method as used for detecting product vectors in $\Im \gr$ (and in $\Ker\gr$), and which is described in Appendix A. The outline of the method is the following. Let $P$ be the orthogonal projection on $\Im\gr$ and $Q$ the orthogonal projection of $\Im\gr^P$. We consider the following bilinear function of the product state $\psi=\phi\otimes\chi$,
\be{bilin}
f(\psi)=\psi^\dag(\iden-P)\psi+\tilde\psi^\dag(\iden-Q)\tilde\psi
\ee
and search for the minima of the function where $f=0$. Since both terms in the expression \pref{bilin} are non-negative, such a solution will give zero for each term separately, and from this follow the relations $P\psi=\psi$ and $Q\tilde\psi=\tilde\psi$, which are equivalent to the two sets of equations \pref{orthocond}. The second term in \pref{bilin} can be rewritten in the following way, $\tilde\psi^\dag(\iden-Q)\tilde\psi=\psi^\dag(\iden-Q^P)\psi$, with $Q^P$ as the partial transpose of $Q$. This gives 
\be{bilin2}
f(\psi)=\psi^\dag(2\iden-P-Q^P)\psi
\ee
and written in this way the function $f(\psi)$ has the same form as the function that is minimized in the search for product vectors in $\Im\gr$ (see Sect.~2.3). The only difference is that the operator $\iden-P$ is replaced by $2\iden-P-Q^P$. The same method to search for product vectors with vanishing value for the function can therefore be used.

The result is that for all the states we have found with $m>N_A+N_B-1$, and where \pref{critlim2} is satisfied as a proper inequality, there are no pairs of conjugate product vectors, and the states are therefore entangled.  In the diagrams these states are located below the dashed green lines. For the states {\em on} the dashed green line in the diagrams the condition \pref{critlim2} is satisfied with equality, and for these states we find a finite number $K$ of pairs of conjugate product vectors that is sufficiently large to satisfy the separability condition \pref{sepcond}. This means that just counting the number of pairs of conjugate product vectors is insufficient to determine if the states are entangled. However, with only a finite number of product vectors available it is possible to check whether the density operator can be reconstructed as a convex combination of these product states. The result is that for all ranks where \pref{critlim2} is satisfied with equality we find entangled states. However in a small number of cases we find both separable and entangled states with the same set of ranks $(m,n)$.  That happens for the $(6,6)$, $(5,7)$ and $(7,5)$ states of the 2x4 system.

\section{Concluding remarks}

The results presented in this paper are based on the use of a numerical method to search for density matrices $\gr$ of a composite, bipartite system with specified values for the ranks of $\gr$ and its partial transpose $\gr^P$. The method works well for systems where the Hilbert space dimension $N$ is not too large, in our calculations with $N\lesssim 20$. In higher dimensions the main problem is a too slow convergence of the iteration procedure. Additional methods to find the number of product vectors in the image and kernel of the density matrices and to determine the dimension of the corresponding face  of the set of PPT states have been described. The results based on the use of these methods are listed in the tables and displayed in the diagrams.

The results obtained for the low-dimensional systems that we have studied reveal several regularities. For sufficiently low ranks we find only separable states of a specific form. Above a certain value for the ranks of $\gr$ and $\gr^P$ the states we find are typically extremal PPT states, until the ranks reach an upper limit. In our discussion we suggest that there are in fact two lower bounds, one for entangled states with full local ranks and another, generally more restrictive, for extremal PPT states with full local ranks. The first one we identify as the bound on entangled PPT states discussed in \cite{Horodecki00}, and we show, by an explicit construction, how this bound can be saturated. Based on certain properties of the extremal states with minimal ranks that are found in our searches, we conjecture a specific value for the second one, the lower bound on the ranks of  extremal PPT states. The property we focus on is the number of product vectors in the image and in the kernel of this state, which makes these states different from the higher rank extremal states. Assuming this property to be present for the lowest rank extremal states in general,  we draw the conclusion about the lower bound.

Above this lower bound we find in our searches a large set of ranks $(m,n)$ of $\gr$ and $\gr^P$ where the states are extremal. There is an upper limit to the ranks of these states, which can be understood as following from a geometrical constraint on the face of the set $\cP$ to which an extremal PPT state belongs. It is of interest to note that all the states we find for sufficiently low ranks  are limited to the symmetric cases $m=n$, whereas for $m$ and $n$ above this limit we find in addition states for essentially all the asymmetric values of the ranks. However, as we have stressed, the states we find by our method should be considered as typical states for the given ranks $(m,n)$. This means that we cannot exclude the presence of untypical states also for ranks where we have not identified any PPT state.

Concerning this last point it is of interest to relate the results for the properties of  the PPT states found in our searches with those of other entangled and extremal PPT states referred to in the literature \cite{Horodecki97,Bennett99, BrussPeres00,DiVincenzo03,Ha03,Clarisse06}. These states are based on special constructions which lead to certain classes of extremal PPT states. Among the states presented by these special constructions we have found no example of states with ranks different from those referred to in our tables and diagrams. The main difference, however, is that these specially constructed states are not necessarily typical in the meaning used here, and in particular the number of product vectors in the image and kernel may be larger than the minimal values that we find in our searches.

In the discussion of our results we have put some emphasis on the properties of the lowest rank extremal PPT states. For all the systems we have studied they are special in the sense that they have no product state in their image, but a complete, finite set of product vectors in their kernel. In a separate publication \cite{LeinaasMyrheim09} we have made a detailed study of these states for the 3x3 system, where they have ranks $(4,4)$. We show there that these states can be related to states
constructed from unextendible product bases,
and the set of such states can be given an explicit parameterization.
We do not know how to generalize this construction to higher
dimensions, and the question of a general parameterization of
extremal PPT states remains as an interesting problem for future work.
\vskip0.5cm

\noi
{\bf\large Acknowledgment}\\
Financial support from the Norwegian Research Council is gratefully acknowledged.

\appendix

\section{The minimum double eigenvalue problem}

Given the Hermitian matrix $A$, the problem considered here is to
minimize the expectation value $\psi^{\dag}A\psi$ over product vectors
$\psi=\phi\otimes\chi$, with the normalization conditions
$\phi^{\dag}\phi=\chi^{\dag}\chi=1$.  We present an iteration method
which may converge to different local minima, depending on the
starting point for the iterations.  We introduce Lagrange multipliers
$\lambda,\mu$ and define
\be{}
f=\sum_{i,j,k,l}
\phi_i^{\ast}\chi_j^{\ast}A_{ij;kl}\phi_k\chi_l
-\lambda\left(\sum_i\phi_i^{\ast}\phi_i-1\right)
-\mu\left(\sum_j\chi_j^{\ast}\chi_j-1\right).
\ee
The following equations must hold at the minimum, or more generally at
an extremum,
\be{}
\frac{\partial f}{\partial\phi_i^{\ast}}
=\sum_{j,k,l}\chi_j^{\ast}A_{ij;kl}\phi_k\chi_l-\lambda\phi_i
=0\;,\qquad
\frac{\partial f}{\partial\chi_j^{\ast}}
=\sum_{i,k,l}
\phi_i^{\ast}A_{ij;kl}\phi_k\chi_l-\mu\chi_j=0\;,
\ee
with
\be{}
\sum_i\phi_i^{\ast}\phi_i=\sum_j\chi_j^{\ast}\chi_j=1\;,
\qquad
\lambda=\mu=\sum_{i,j,k,l}
\phi_i^{\ast}\chi_j^{\ast}A_{ij;kl}\phi_k\chi_l\;.
\ee

Given an approximate solution $\psi=\phi\otimes\chi$, we compute a
next approximation $\psi'=\phi'\otimes\chi'$ as follows.  We compute
$x=B\phi-\lambda\phi$, $y=C\chi-\lambda\chi$, with
$\lambda=\psi^{\dag}A\psi$ and
\be{}
B_{ik}=\sum_{j,l}\chi_j^{\ast}A_{ij;kl}\chi_l\;,
\qquad
C_{jl}=\sum_{i,k}\phi_i^{\ast}A_{ij;kl}\phi_k\;.
\ee
In practice, we compute $z=A\psi$,
$u_i=\sum_j\chi_j^{\ast}z_{ij}$,
$v_j=\sum_i\phi_i^{\ast}z_{ij}$,
$\lambda=\phi^{\dag}u=\chi^{\dag}v$,
$x=u-\lambda\phi$,
$y=v-\lambda\chi$.
Then $\phi'=N_1(\phi+\epsilon x)$,
$\chi'=N_2(\chi+\epsilon y)$,
where $N_1,N_2$ are normalization factors, and where $\epsilon$
is determined in the following way.
To first order in $\epsilon$ we have
\be{}
(\phi+\epsilon x)\otimes(\chi+\epsilon y)=\psi+\epsilon w\;,
\ee
with $w=\phi\otimes y+x\otimes\chi$.  Note that
$\phi^{\dag}x=\chi^{\dag}y=0$, and hence $\psi^{\dag}w=0$.  We compute
$\epsilon$ by minimizing $s^{\dag}As/s^{\dag}s$ with
$s=\psi+\epsilon w$.  This is an eigenvalue problem in the two
dimensional subspace spanned by $\psi$ and $w$, and it can be solved
analytically.

%%%%%%%%%%
\newpage
\section{Tables}

\begin{table}[h!]
\centering
\scalebox{0.85}{%
\begin{tabular}{|cccccc|}
\multicolumn{6}{c}{\bf \Large 2x4}\\
\hline
\multicolumn{1}{|c|}{\textbf{$\mathbf{(m,n)}$}} & \multicolumn{1}{c|}{$\mathbf{m^2 + n^2 - N^2}$} & \multicolumn{1}{c|}{\textbf{dim}\,{\boldmath$\cal F$}} & \multicolumn{1}{c|}{$\mathbf{(r_A,r_B)}$} & \multicolumn{1}{c|}{\textbf{\#pv [Im\,$\rho$]}} & \multicolumn{1}{c|}{\textbf{\#pv  [Ker\,$\rho$]}} 
\\
\hline\hline
(8,8) & 64 & 64 & (2,4) & $\infty$/8 & 0 \\ \hline
(8,7) & 49 & 49 & (2,4) & $\infty$/8 & 0 \\ \hline
(8,6) & 36 & 36 & (2,4) & $\infty$/8 & 0 \\ \hline
(7,7) & 34 & 34 & (2,4) & $\infty$/7 & 0 \\ \hline
(8,5) & 25 & 25 & (2,4) & $\infty$/8 & 0 \\ \hline % NY
(7,6) & 21 & 21 & (2,4) & $\infty$/7 & 0 \\ \hline
(7,5) & 10 & 10 & (2,4) & $\infty$/7 & 0 \\ \hline
(6,6) & 8 & 8 & (2,4) & $\infty$/6 & 0 \\ \hline
(6,5) & -3 & 1 & (2,4) & $\infty$/6 & 0 \\ \hline
(5,5) & -14 & 1 & (2,4) & $\infty$/5 & 0 \\ \hline
(4,4) & -32 & 4 & (2,4) & 4/4 & 4/4 \\ \hline
(3,3) & -46 & 3 & (2,3) & 3/3 & $\infty$/5 \\ \hline
(2,2) & -56 & 2 & (2,2) & 2/2 & $\infty$/6 \\ \hline
(1,1) & -62 & 1 & (1,1) & 1/1 & $\infty$/8 \\ \hline
\multicolumn{6}{c}{ }\\
\multicolumn{6}{c}{ }\\
\multicolumn{6}{c}{\bf \Large 2x5}\\
\hline
\multicolumn{1}{|c|}{\textbf{$\mathbf{(m,n)}$}} & \multicolumn{1}{c|}{$\mathbf{m^2 + n^2 - N^2}$} & \multicolumn{1}{c|}{\textbf{dim}\,{\boldmath$\cal F$}} & \multicolumn{1}{c|}{$\mathbf{(r_A,r_B)}$} & \multicolumn{1}{c|}{\textbf{\#pv [Im\,$\rho$]}} & \multicolumn{1}{c|}{\textbf{\#pv  [Ker\,$\rho$]}} 
\\
\hline\hline
(10,10) & 100 & 100 & (2,5) & $\infty$/10 & 0 \\ \hline
(10,9) & 81 & 81 & (2,5) & $\infty$/10 & 0 \\ \hline
(10,8) & 64 & 64 & (2,5) & $\infty$/10 & 0  \\ \hline
(9,9) & 62 & 62 & (2,5) & $\infty$/9 & 0 \\ \hline
(10,7) & 49 & 49 & (2,5) & $\infty$/10 & 0 \\ \hline %NY
(9,8) & 45 & 45 & (2,5) & $\infty$/9 & 0 \\ \hline
(9,7) & 30 & 30 & (2,5) & $\infty$/9 & 0  \\ \hline
(8,8) & 28 & 28 & (2,5) & $\infty$/8 & 0 \\ \hline
(9,6) & 17& 17 & (2,5) & $\infty$/9 & 0 \\ \hline
(8,7) & 13 & 13 & (2,5) & $\infty$/8 & 0 \\ \hline
(8,6) & 0 & 1 & (2,5) & $\infty$/8 & 0 \\ \hline
(7,7) & -2 & 1 & (2,5) & $\infty$/7 & 0 \\ \hline
(7,6) & -15 & 1 & (2,5) & $\infty$/7 & 0 \\ \hline
(6,6) & -28 & 1 & (2,5) & $\infty$/6 & 0 \\ \hline
(5,5) & -50 & 5 & (2,5) & 5/5 & 5/5 \\ \hline
(4,4) & -68 & 4 & (2,4) & 4/4 & $\infty$/6 \\ \hline
(3,3) & -82 & 3 & (2,3) & 3/3 & $\infty$/7 \\ \hline
(2,2) & -92 & 2 & (2,2) & 2/2 & $\infty$/8 \\ \hline
(1,1) & -98 & 1 & (1,1) & 1/1 & $\infty$/9 \\ \hline
\end{tabular}}
\caption{ Numerical results for the 2x4 and the 2x5 systems. The first column lists the ranks of $\gr$ and $\gr^P$ where PPT states have been found. The second column lists the lower limit for the value of the dimension of the face of $\cK(\cP)$ for the given ranks $(m,n)$, while the third column lists the actual values of the dimensions for the states we have found. The fourth column lists the values of the local ranks with respect to subsystems $A$ and $B$. The fifth and sixth columns give the number of product vectors in $\Im\gr$ and $\Ker\gr$, respectively. In each of these columns two numbers are given, with the number to the left as the total number and the one to the right as the number of linearly independent product vectors. $\infty$ indicates that we find no upper limit to the number of product vectors that can be generated. In  the present tables the extremal PPT states with no product state in $\Im\gr$ and a complete set in $\Ker\gr$, which we find in all the other tables, are missing.  } \label{tab2N}
\end{table}
%%%%%%%%%%
%\clearpage
\begin{table}[h!]
\centering
\scalebox{0.85}{%
\begin{tabular}{|cccccc|}
\multicolumn{6}{c}{\bf \Large 3x3}\\
\hline
\multicolumn{1}{|c|}{\textbf{$\mathbf{(m,n)}$}} & \multicolumn{1}{c|}{$\mathbf{m^2 + n^2 - N^2}$} & \multicolumn{1}{c|}{\textbf{dim}\,{\boldmath$\cal F$}} & \multicolumn{1}{c|}{$\mathbf{(r_A,r_B)}$} & \multicolumn{1}{c|}{\textbf{\#pv [Im\,$\rho$]}} & \multicolumn{1}{c|}{\textbf{\#pv  [Ker\,$\rho$]}} 
\\
\hline\hline
(9,9) & 81 & 81 & (3,3) & $\infty$/9 & 0 \\ \hline
(9,8) & 64 & 64 & (3,3) & $\infty$/9 & 0 \\ \hline
(9,7) & 49 & 49 & (3,3) & $\infty$/9 & 0 \\ \hline
(8,8) & 47 & 47 & (3,3) & $\infty$/8 & 0 \\ \hline
(9,6) & 36 & 36 & (3,3) & $\infty$/9 & 0 \\ \hline % NY
(8,7) & 32 & 32 & (3,3) & $\infty$/8 & 0 \\ \hline
(8,6) &  19 & 19 & (3,3) & $\infty$/8 & 0 \\ \hline
(7,7) &  17 & 17 & (3,3) & $\infty$/7 & 0 \\ \hline
(8,5) &    8 &    8 & (3,3) & $\infty$/8 & 0 \\ \hline
(7,6) &    4 &    4 & (3,3) & $\infty$/7 & 0 \\ \hline
(7,5) &   -7 &    1 & (3,3) & $\infty$/7 & 0 \\ \hline
(6,6) &   -9 &    1 & (3,3) & $\infty$/6 & 0 \\ \hline
(6,5) & -20 &    1 & (3,3) & $\infty$/6 & 0 \\ \hline 
(5,5) & -31 &    1 & (3,3) & 6/5 & 0 \\ \hline
(4,4) & -49 &    1 & (3,3) & 0 & 6/5 \\ \hline
(3,3) & -63 &    3 & (3,3) & 3/3 & $\infty$/6 \\ \hline
(2,2) & -73 &    2 & (2,2) & 2/2 & $\infty$/7 \\ \hline
(1,1) & -79 &    1 & (1,1) & 1/1 & $\infty$/8 \\ \hline
\multicolumn{6}{c}{ }\\
\multicolumn{6}{c}{ }\\
\multicolumn{6}{c}{\bf \Large 3x4}\\
\hline
\multicolumn{1}{|c|}{\textbf{$\mathbf{(m,n)}$}} & \multicolumn{1}{c|}{$\mathbf{m^2 + n^2 - N^2}$} & \multicolumn{1}{c|}{\textbf{dim}\,{\boldmath$\cal F$}} & \multicolumn{1}{c|}{$\mathbf{(r_A,r_B)}$} & \multicolumn{1}{c|}{\textbf{\#pv [Im\,$\rho$]}} & \multicolumn{1}{c|}{\textbf{\#pv  [Ker\,$\rho$]}} 
\\
\hline\hline
12,11) & 121 & 121 & (3,4) & $\infty$/12 & 0 \\ \hline
(12,10) & 100 & 100 & (3,4) & $\infty$/12 & 0  \\ \hline
(11,11) & 98 & 98 & (3,4) & $\infty$/11 & 0 \\ \hline
(12,9) & 81 & 81 & (3,4) & $\infty$/12 & 0 \\ \hline
(11,10) & 77 & 77 & (3,4) & $\infty$/11 & 0 \\ \hline
%\textcolor{red}{(12,8)} 
(12,8) & 64 & 64 & (3,4) & $\infty$/12 & 0 \\ \hline % NY
(11,9) & 58 & 58 & (3,4) & $\infty$/11 & 0  \\ \hline
(10,10) & 56 & 56 & (3,4) & $\infty$/10 & 0 \\ \hline
(11,8) & 41& 41 & (3,4) & $\infty$/11 & 0\\ \hline
(10,9) & 37 & 37 & (3,4) & $\infty$/10 & 0 \\ \hline
(11,7) & 26 & 26 & (3,4) & $\infty$/11 & 0 \\ \hline
(10,8) & 20 & 20 & (3,4) & $\infty$/10 & 0 \\ \hline
(9,9) & 18 & 18 & (3,4) & $\infty$/9 & 0 \\ \hline
%\textcolor{red}{(11,6)} 
(11,6) & 13 & 13 & (3,4) & $\infty$/11 & 0 \\ \hline % NY
(10,7) & 5 & 5 & (3,4) & $\infty$/10 & 0 \\ \hline
(9,8) & 1 & 1 & (3,4) & $\infty$/9 & 0 \\ \hline
(10,6) & -8 & 1 & (3,4) & $\infty$/10 & 0 \\ \hline
(9,7) & -14 & 1 & (3,4) & $\infty$/9 & 0 \\ \hline
(8,8) & -16 & 1 & (3,4) & $\infty$/8 & 0 \\ \hline
(9,6) & -27 & 1 & (3,4) & $\infty$/9 & 0 \\ \hline
(8,7) & -31 & 1 & (3,4) & $\infty$/8 & 0 \\ \hline
(8,6) & -44 & 1 & (3,4) & $\infty$/8 & 0 \\ \hline
(7,7) & -46 & 1 & (3,4) & 10/7 & 0 \\ \hline
(7,6) & -59 & 1 & (3,4) & 10/7 & 0 \\ \hline
(6,6) & -72 & 1 & (3,4) & 0 & 0 \\ \hline
(5,5) & -94 & 1 & (3,4) & 0 & 10/7 \\ \hline
\multirow{2}{*}{(4,4)} & \multirow{2}{*}{-112} & 1 & (3,3) & 0 & $\infty$/8 \\
 & & 4 & (3,4) & 4/4 & $\infty$/8 \\ \hline
(3,3) & -126  & 3 & (3,3) & 3/3 & $\infty$/9 \\ \hline
(2,2) & - 136 & 2 & (2,2) & 2/2 & $\infty$/10 \\ \hline
(1,1) & -142 & 1 & (1,1) & 1/1 & $\infty$/11 \\ \hline
\end{tabular}}
\caption{Numerical results for the 3x3 and the 3x4 systems. For explanations see Table 1.  Except for one case in the 3x4 system we find only one type of state, characterized by the listed properties, for each set of ranks $(m,n)$. The exception is the $(4,4)$ where we find both extremal PPT states with less than full local ranks and separable states of the same construction a those of lower ranks. } \label{tab3N}
\end{table}
%%%%%%%%%%
\clearpage
\begin{table}[h!]
\centering
\scalebox{0.85}{%
%\begin{supertabular}{|cccccc|}
\begin{tabular}{|cccccc|}
\multicolumn{6}{c}{\bf \Large 3x5}\\
\hline
\multicolumn{1}{|c|}{\textbf{$\mathbf{(m,n)}$}} & \multicolumn{1}{c|}{$\mathbf{m^2 + n^2 - N^2}$} & \multicolumn{1}{c|}{\textbf{dim}\,{\boldmath$\cal F$}} & \multicolumn{1}{c|}{$\mathbf{(r_A,r_B)}$} & \multicolumn{1}{c|}{\textbf{\#pv [Im\,$\rho$]}} & \multicolumn{1}{c|}{\textbf{\#pv  [Ker\,$\rho$]}} 
\\
\hline\hline
(15,14) & 196 & 196 & (3,5) & $\infty$/15 & 0 \\ \hline
(15,13) & 169 & 169 & (3,5) & $\infty$/15 & 0 \\\hline
(14,14) & 167 & 167 & (3,5) & $\infty$/14 & 0 \\\hline
(15,12) & 144 & 144 & (3,5) & $\infty$/15 & 0 \\\hline
(14,13) & 140 & 140 & (3,5) & $\infty$/14 & 0 \\\hline
(15,11) & 121 & 121 & (3,5) & $\infty$/15 & 0 \\\hline
(14,12) & 115 & 115 & (3,5) & $\infty$/14 & 0 \\\hline
(13,13) & 113 & 113 & (3,5) & $\infty$/13 & 0 \\ \hline
(14,11) & 92 & 92 & (3,5) & $\infty$/14 & 0 \\\hline
(13,12) & 88 & 88 & (3,5) & $\infty$/13 & 0 \\\hline
(14,10) & 71   & 71    & (3,5) & $\infty$/14 & 0 \\\hline
(13,11) & 65 & 65 & (3,5) & $\infty$/13 & 0 \\\hline
(12,12)   & 63 & 63      & (3,5) & $\infty$/12 & 0 \\\hline
(14,9) & 52 & 52 & (3,5) & $\infty$/14 & 0 \\\hline
(13,10) & 44   & 44    & (3,5) & $\infty$/13 & 0 \\\hline
(12,11)   & 40   & 40         & (3,5) & $\infty$/12 & 0 \\\hline
(14,8) & 35 & 35 & (3,5) & $\infty$/14 & 0 \\\hline
(13,9) & 25 & 25      & (3,5) & $\infty$/13 & 0 \\\hline
(14,7) & 20 & 20 & (3,5) & $\infty$/14 & 0 \\\hline
(12,10) & 19 & 19      & (3,5) & $\infty$/12 & 0 \\\hline
(11,11) & 17 & 17      & (3,5) & $\infty$/11 & 0 \\\hline
(13,8) & 8 & 8      & (3,5) & $\infty$/13 & 0 \\\hline
(12,9) & 0 & 1      & (3,5) & $\infty$/12 & 0 \\\hline
(11,10) & -4    & 1         & (3,5) & $\infty$/11 & 0 \\\hline
(13,7) & -7 & 1 & (3,5) & $\infty$/13 & 0 \\\hline
(12,8) & -17 & 1      & (3,5) & $\infty$/12 & 0 \\\hline
(11,9) & -23 & 1          & (3,5) & $\infty$/11 & 0 \\\hline
(10,10) & -25 & 1 & (3,5) & $\infty$/10 & 0 \\ \hline
(12,7) & -32 & 1 & (3,5) & $\infty$/12 & 0 \\ \hline % NY
(11,8) & -40 & 1 & (3,5) & $\infty$/11 & 0 \\ \hline
(10,9) & -44 & 1 & (3,5) & $\infty$/10 & 0 \\\hline
(11,7) & -55 & 1 & (3,5) & $\infty$/11 & 0 \\ \hline
(10,8) & -61 & 1 & (3,5) & $\infty$/10 & 0 \\\hline
(9,9) & -63 & 1 & (3,5) & 15/9 & 0 \\\hline
(10,7) & -76 & 1 & (3,5) & $\infty$/10 & 0 \\\hline
(9,8) & -80 & 1 & (3,5) & 15/9 & 0 \\\hline
(9,7) & -95 & 1 & (3,5) & 15/9 & 0 \\\hline
(8,8) & -97 & 1 & (3,5) & 0 & 0 \\\hline
(8,7) & -112 & 1 & (3,5) & 0 & 0 \\\hline
(7,7) & -127 & 1 & (3,5) & 0 & 0 \\\hline
(6,6) & -153 & 1 & (3,5) & 0 & 15/9 \\\hline
\multirow{2}{*}{(5,5)} & \multirow{2}{*}{-175} & 1 & (3,4) & 0 & $\infty$/10 \\
& & 5 & (3,5) & 5/5 & $\infty$/10 \\\hline
(4,4) & -193 & 4 & (3,4) & 4/4 & $\infty$/11 \\\hline
(3,3) & -207 & 3 & (3,3) & 3/3 & $\infty$/12 \\\hline
(2,2) & -217 & 2 & (2,2) & 2/2 & $\infty$/13 \\\hline
(1,1) & -223 & 1 & (1,1) & 1/1 & $\infty$/14 \\\hline
%\end{supertabular}}
\end{tabular}}
\caption{Numerical results for the 3x5 systems. For explanations see Table 1.  We here find two types of states with ranks $(5,5)$.} \label{tab35}
\end{table}

%\clearpage
%%%%%%%%%%%
\begin{table}[h]
\centering
% HER KAN DU SKALERE STØRRELSEN PÅ TABELLEN MED SCALEBOX. TALLET I { } ANGIR
% STØRRELSE I PROSENT AV ORIGINAL STØRRELSE. 1 FOR 100% OG 0.1 FOR 10% ETC.
\scalebox{0.80}{%
\begin{tabular}{|cccccc|c|}
\multicolumn{6}{c}{\bf \Large 4x4}\\
\hline
\multicolumn{1}{|c|}{\textbf{$\mathbf{(m,n)}$}} & \multicolumn{1}{c|}{$\mathbf{m^2 + n^2 - N^2}$} & \multicolumn{1}{c|}{\textbf{dim}\,{\boldmath$\cal F$}} & \multicolumn{1}{c|}{$\mathbf{(r_A,r_B)}$} & \multicolumn{1}{c|}{\textbf{\#pv [Im\,$\rho$]}} & \multicolumn{1}{c|}{\textbf{\#pv  [Ker\,$\rho$]}} 
\\
\hline\hline
(16,16) & 256 & 256 & (4,4) & $\infty$/16 & 0   \\ \hline
(16,15) & 225 & 225 & (4,4) & $\infty$/16 & 0   \\ \hline
(16,14) & 196 & 196 & (4,4) & $\infty$/16 & 0   \\\hline
(15,15) & 194 & 194 & (4,4) & $\infty$/15 & 0   \\\hline
(16,13) & 169 & 169 & (4,4) & $\infty$/16 & 0   \\\hline
(15,14) & 165 & 165 & (4,4) & $\infty$/15 & 0   \\\hline
{(16,12)} & 144 & 144 & (4,4) & $\infty$/16 & 0   \\\hline % NY
(15,13) & 138 & 138 & (4,4) & $\infty$/15 & 0   \\\hline
(14,14) & 136 & 136 & (4,4) & $\infty$/14 & 0   \\\hline
{(16,11)} & 121 & 121 & (4,4) & $\infty$/16 & 0   \\\hline % NY
(15,12) & 113 & 113 & (4,4) & $\infty$/15 & 0   \\\hline
(14,13) & 109 & 109 & (4,4) & $\infty$/14 & 0   \\\hline
(15,11) & 90   & 90    & (4,4) & $\infty$/15 & 0   \\\hline
(14,12) & 84   & 84    & (4,4) & $\infty$/14 & 0   \\\hline
(13,13) & 82 & 82      & (4,4) & $\infty$/13 & 0   \\\hline
(15,10) & 69 & 69      & (4,4) & $\infty$/15 & 0   \\ \hline
(14,11) & 61 & 61      & (4,4) & $\infty$/14 & 0   \\\hline
(13,12) & 57 & 57      & (4,4) & $\infty$/13 & 0   \\\hline
{(15,9)} & 50 & 50 & (4,4) & $\infty$/15 & 0   \\\hline % NY
(14,10) & 40 & 40      & (4,4) & $\infty$/14 & 0   \\\hline
{(15,8)} & 33 & 33 & (4,4) & $\infty$/15 & 0   \\\hline % NY
(13,11) & 34 & 34      & (4,4) & $\infty$/13 & 0   \\\hline
(12,12) & 32 & 32      & (4,4) & $\infty$/12 & 0   \\ \hline
(14,9)   & 21 & 21      & (4,4) & $\infty$/14 & 0   \\\hline
{(15,7)}   & 18 & 18     & (4,4) & $\infty$/15 & 0   \\\hline % NY
(13,10) & 13 & 13      & (4,4) & $\infty$/13 & 0   \\\hline
(12,11) & 9 & 9           & (4,4) & $\infty$/12 & 0   \\ \hline
(14,8)   & 4   & 4         & (4,4) & $\infty$/14 & 0   \\\hline
(13,9) & -6    & 1         & (4,4) & $\infty$/13 & 0   \\\hline
{(14,7)}   & -11 & 1     & (4,4) & $\infty$/14 & 0   \\\hline % NY
(12,10) & -12 & 1 & (4,4) & $\infty$/12 & 0  \\\hline
(11,11) & -14 & 1 & (4,4) & $\infty$/11 & 0   \\\hline
(13,8) & -23 & 1          & (4,4) & $\infty$/13 & 0   \\\hline
(12,9) & -31 & 1 & (4,4) & $\infty$/12 & 0   \\\hline
(11,10) & -35 & 1 & (4,4) & $\infty$/11 & 0   \\\hline
{(13,7)}   & -38 & 1     & (4,4) & $\infty$/13 & 0   \\\hline % NY
(12,8) & -48 & 1 & (4,4) & $\infty$/12 & 0   \\\hline
(11,9) & -54 & 1 & (4,4) & $\infty$/11 & 0   \\\hline
(10,10) & -56 & 1 & (4,4) & 20/10 & 0   \\\hline
(12,7) & -63 & 1 & (4,4) & $\infty$/12 & 0 \\\hline %NY
(11,8) & -71 & 1 & (4,4) & $\infty$/11 & 0   \\\hline
(10,9) & -75 & 1 & (4,4) & 20/10 & 0   \\\hline
(11,7) & -86 & 1 & (4,4) & $\infty$/11 & 0 \\\hline % NY
(10,8) & -92 & 1 & (4,4) & 20/10 & 0   \\\hline
(9,9) & -94 & 1 & (4,4) & 0 & 0   \\\hline
(10,7) & -107 & 1 & (4,4) & 20/10 & 0   \\\hline
(9,8) & -111 & 1 & (4,4) & 0 & 0   \\\hline
(9,7) & -126 & 1 & (4,4) & 0 & 0   \\\hline
(8,8) & -128 & 1 & (4,4) & 0 & 0   \\\hline
(8,7) & -143 & 1 & (4,4) & 0 & 0   \\\hline
(7,7) & -158 & 1 & (4,4) & 0 & 0   \\\hline
(6,6) & -184 & 1 & (4,4) & 0 & 20/10   \\\hline
\multirow{3}{*}{(5,5)} & \multirow{3}{*}{-206} & 1 & (4,3) & 0 & $\infty$/11   \\
& & 2 & (4,4) & 1/1  & $\infty$/11   \\
& & 5 &(4,4)  & 5/5  & $\infty$/11   \\\hline
\multirow{2}{*}{(4,4)} & \multirow{2}{*}{-224} & 1 & (3,3) & 0 & $\infty$/12   \\
& & 4 & (4,4) & 4/4 & $\infty$/12   \\\hline
(3,3) & -238 & 3 & (3,3) & 3/3 & $\infty$/13   \\\hline
(2,2) & -248 & 2 & (2,2) & 2/2 & $\infty$/14   \\\hline
(1,1) & -254 & 1 & (1,1) & 1/1 & $\infty$/15   \\\hline
\end{tabular}}
\caption{Numerical results for the 4x4 systems. For explanations see Table 1. The pattern of the listed properties is much like that of the other tables, but here with more low-rank states below the lowest-rank ($m=n=6$) extremal PPT state with full local ranks.  } \label{tab44}
\end{table}

\clearpage
%\newpage
%%%%%%%%%

\end{document}